\documentclass[pdflatex,sn-mathphys-num]{sn-jnl}

\usepackage{graphicx}
\usepackage{multirow}
\usepackage{amsmath,amssymb,amsfonts}
\usepackage{bm}
\usepackage{booktabs}
\usepackage{tabularx}
\usepackage{array}
\usepackage{xcolor}
\usepackage{dcolumn}

\newcommand{\rvadd}[1]{#1}
\newcommand{\rvdel}[1]{}
\newcommand{\rvmdel}[1]{}
\newcommand{\rvdelhead}[1]{}
\newcommand{\rvdelblock}[1]{}
\newcommand{\rvtodo}[1]{}
\newcommand{\rvnote}[1]{}

\newcommand{\sddel}[1]{}
\newcommand{\sdmdel}[1]{}
\newcommand{\sddelblock}[1]{}
\newcommand{\sdtodo}[1]{}
\newcommand{\sdnote}[1]{}

\newcommand{\onlinecite}[1]{\cite{#1}}

\begin{document}

\title[Effective transport physics and optimal feedback control]{Identification of effective transport physics and optimal feedback control: Experimental demonstration on a 1600-fan wind generator}

\author*[1,2]{\fnm{T.} \sur{Shaqarin}}\email{tshagareen@ttu.edu.jo}
\author[3,4]{\fnm{I.} \sur{Al-Manasra}}
\author[5]{\fnm{T.} \sur{Zhang}}
\author[6]{\fnm{A.} \sur{Ianiro}}
\author[6]{\fnm{S.} \sur{Discetti}}
\author[7,8,9]{\fnm{B. R.} \sur{Noack}}\email{bernd.noack@szu.edu.cn}

\affil*[1]{\orgdiv{Department of Robotics and Artificial Intelligence}, \orgname{American University of Madaba}, \orgaddress{\city{Madaba}, \country{Jordan}}}
\affil[2]{\orgdiv{Department of Mechanical Engineering}, \orgname{Tafila Technical University}, \orgaddress{\city{Tafila}, \postcode{66110}, \country{Jordan}}}
\affil[3]{\orgdiv{Department of Mechanical Engineering}, \orgname{Al Hussein Technical University}, \orgaddress{\city{Amman}, \country{Jordan}}}
\affil[4]{\orgdiv{Department of Mechanical Engineering}, \orgname{University of Jordan}, \orgaddress{\city{Amman}, \country{Jordan}}}
\affil[5]{\orgname{Ningbo Dongteng Low-Altitude Technology Co., Ltd.}, \orgaddress{\city{Ningbo}, \postcode{315200}, \country{People's Republic of China}}}
\affil[6]{\orgdiv{Department of Aerospace Engineering}, \orgname{Universidad Carlos III de Madrid}, \orgaddress{\street{Av. de la Universidad 30}, \postcode{28911}, \city{Legan\'es}, \country{Spain}}}
\affil[7]{\orgdiv{College of Mechatronics and Control Engineering}, \orgname{Shenzhen University}, \orgaddress{\city{Shenzhen}, \postcode{518060}, \country{People's Republic of China}}}
\affil[8]{\orgdiv{College of Urban Transport and Logistics}, \orgname{Shenzhen Technology University}, \orgaddress{\city{Shenzhen}, \postcode{518118}, \country{People's Republic of China}}}
\affil[9]{\orgdiv{Low-Altitude Economy Science and Technology Research Institute}, \orgname{Harbin Institute of Technology Shenzhen}, \orgaddress{\city{Shenzhen}, \postcode{518055}, \country{People's Republic of China}}}

\abstract{Distributed-input distributed-output (DIDO) systems arise in numerous
engineering applications where large numbers of actuators generate
spatially distributed physical fields that are monitored by distributed sensors. Developing physically
interpretable models that enable real-time optimal control of these systems remains
challenging because of the high dimensionality and strong spatial
coupling between inputs and outputs. In this work, we propose a physics-based
framework for modeling and closed-loop optimal control of
DIDO systems from experimental input--output data. First, an effective partial differential
equation (PDE) is selected among candidate transport models and
discretized on the actuator grid to obtain a minimum-grid representation. Second, the PDE is solved in closed form, which yields an input--output operator with a four physically interpretable parameters. Third, the operator is incorporated into an iterative, constrained quadratic program that accounts for the actuator limits. 
The framework is demonstrated on the inner $40\times40$ section of the world's highest-resolution Fan Array Wind Generator (FAWG), comprising 1600 independently controlled fans with 36 distributed velocity measurements. The model describes the spatial interaction between the 1600 actuators and the measured velocity field with normalized mean absolute prediction errors of 11--20\% for representative actuation patterns. The closed-loop controller reproduces all three prescribed two-dimensional
velocity fields with a normalized mean absolute error below
$2.3\%$ in fewer than 10 iterations.
%\sddel{ We expect the identified PDE model to become a foundational pillar of FAWG control, as both its structure and the order of magnitude of the diffusion coefficients are consistent with first-principles shear turbulence approximations.}
}


\maketitle

% Requires:
% \usepackage{tabularx}

\section*{NOMENCLATURE}

\noindent\textbf{Dimensions and indices}

\begin{tabularx}{\columnwidth}{@{}l@{\hspace{1em}}X@{}}
$N_b$       & Number of independently controlled actuators \\
$N_s$       & Number of sensors (measured plant outputs) \\
$N_m$       & Number of command--measurement experiments \\
$N_x,N_y$ & Actuator-matrix dimensions in the $x$ and $y$ directions \\
$N_g$       & Total number of spatial grid points, $N_g=N_xN_y$ \\
$k$         & Closed-loop iteration index
\end{tabularx}

\vspace{0.5em}
\noindent\textbf{Experimental input--output quantities}

\begin{tabularx}{\columnwidth}{@{}l@{\hspace{1em}}X@{}}
$\mathbf{b}$          & Distributed actuator-command vector,
                        $\mathbf{b}\in\mathbb{R}^{N_b}$ \\
$b_{\max}$            & Upper bound on the actuator commands \\
$\mathbf{s}$          & Measured output (velocity) vector,
                        $\mathbf{s}\in\mathbb{R}^{N_s}$ \\
$\mathbf{s}_d$        & Desired output (target velocity) vector \\
$\mathbf{e}$          & Output tracking-error vector,
                        $\mathbf{e}=\mathbf{s}_d-\mathbf{s}$ \\
$\Delta\mathbf{b}$    & Incremental actuator-command vector \\
$\mathbf{B}$          & Experimental actuator-command data matrix,
                        $\mathbf{B}\in\mathbb{R}^{N_b\times N_m}$ \\
$\mathbf{S}$          & Experimental output data matrix,
                        $\mathbf{S}\in\mathbb{R}^{N_s\times N_m}$ \\
$\mathbf{G}$          & Empirical input--output operator,
                        $\mathbf{G}=\mathbf{S}\mathbf{B}^{\dagger}$ \\
$(\cdot)^\dagger$     & Moore--Penrose pseudoinverse \\
\end{tabularx}

\vspace{0.5em}
\noindent\textbf{Continuous transport model}

\begin{tabularx}{\columnwidth}{@{}l@{\hspace{1em}}X@{}}
$x,y$        & Transverse spatial coordinates \\
$z$          & Streamwise coordinate \\
$Z$ & Dimensionless streamwise evolution coordinate, $Z=z/W$; measurement plane at $Z=1$ \\
$W$ & Width of the active FAWG section ($3.25~\mathrm{m}$) \\
$U(x,y,Z)$   & Continuous streamwise velocity field \\
$b(x,y)$      & Spatially distributed actuator-command field \\
$\nu_x,\nu_y$ & Effective transverse diffusion coefficients in the $x$ and $y$ directions \\
$\nu_{t,x},\,\nu_{t,y}$ & Effective transverse eddy viscosities in the $x$ and $y$ directions \\
$v_x,v_y$ & Effective transverse advection coefficients (candidate models) \\
$U_c$ & Characteristic convection velocity \\
$F_b$ & Streamwise body force per unit mass \\
$u_m,\,l_m$ & Eddy velocity and mixing length \\
$\beta$      & Effective linear decay coefficient \\
$\gamma$     & Effective actuator forcing gain \\
$\boldsymbol{\theta}$ & Physical-parameter vector,
               $\boldsymbol{\theta}
               =[\nu_x,\nu_y,\beta,\gamma]^T$ \\
\end{tabularx}

\vspace{0.5em}
\noindent\textbf{Discrete physics-based model}

\begin{tabularx}{\columnwidth}{@{}l@{\hspace{1em}}X@{}}
$\mathbf{U}(Z)$       & Discrete velocity-field state vector on the actuator grid \\
$\mathbf{T}_x,\mathbf{T}_y$
                      & One-dimensional finite-difference second-derivative
                        operators in the $x$ and $y$ directions \\
$\mathbf{L}_x,\mathbf{L}_y$
                      & Two-dimensional discrete Laplacian operators in the
                        $x$ and $y$ directions \\
$\mathbf{I}$          & Identity matrix of appropriate dimension \\
$\otimes$             & Kronecker product \\
$\mathbf{A}_{\boldsymbol{\theta}}$
                      & Discrete PDE evolution operator \\
$\mathbf{C}$          & Grid-to-sensor interpolation operator,
                        $\mathbf{C}\in\mathbb{R}^{N_s\times N_g}$ \\
$\mathbf{H}_{\boldsymbol{\theta}}$
                      & Physics-based analytical input--output operator,
                        $\mathbf{H}_{\boldsymbol{\theta}}
                        \in\mathbb{R}^{N_s\times N_b}$ 
\end{tabularx}

\vspace{0.5em}
\noindent\textbf{Optimal control}

\begin{tabularx}{\columnwidth}{@{}l@{\hspace{1em}}X@{}}
$\lambda$             & Control-effort regularization parameter \\
$\mathbf{b}_k$        & Actuator-command vector at closed-loop iteration $k$ \\
$\mathbf{s}_k$        & Measured output vector at closed-loop iteration $k$ \\
$\mathbf{e}_k$        & Tracking-error vector at iteration $k$,
                        $\mathbf{e}_k=\mathbf{s}_d-\mathbf{s}_k$ \\
$\Delta\mathbf{b}_k$  & Optimal incremental actuator-command vector at iteration $k$ 

\end{tabularx}

% --- INTRODUCTION SECTION (NO HEADER) ---
%% =====================================================================
\section{Introduction}\label{sec:intro}
%% =====================================================================
%\textbf{Distributed systems}
Spatially distributed transport phenomena arise throughout nature and engineering. Examples include turbulent flows, ocean circulation, thermal transport, reactive flows, and biological transport processes. Distributed-parameter systems with distributed sensing and actuation,
here referred to as distributed-input distributed-output (DIDO) systems,
appear in many engineering applications where distributed actuators are employed to generate or manipulate spatially varying physical fields, such as arrays of jets or other flow actuators in active flow control, distributed heating in thermal and chemical processes, turbine arrays in wind farms, and fan-array wind generators, the application considered here.  
The large number of actuators and the interaction between neighboring actuators result in highly coupled input-output dynamics, making accurate modeling and real-time control challenging \cite{xu2025advances,gehring2023control, zhao2024dual,sun2025learning}.In many of these systems the coupling is caused by transport: the influence of each actuator is carried towards the sensors and spread transversally by convection and diffusion. The input--output map is then shaped by a few transport mechanisms, which suggests describing it by an effective transport equation with a small number of parameters.

Existing modeling approaches largely fall into three categories.
Physics-based models derive the governing equations from first principles and provide valuable physical
insight. However, obtaining an accurate reduced-order input-output model
for complex experimental systems is often difficult and usually requires
problem-dependent assumptions \cite{brunton2015closed,rowley2017model,benner2015survey}. Data-driven models \cite{allen2025end,stefan2026data,shaqarin2025kernel} learn the
input-output relationship directly from experimental measurements
without requiring prior knowledge of the governing physics. Although
they can reproduce the measured response, they typically require extensive training data and computational effort, and provide limited physical insight
into the underlying transport mechanisms. Equation-discovery methods \cite{schmidt2009distilling,brunton2016discovering,Rudy2017,yuan2019data} are an exception: they yield interpretable governing equations but require the state to be sampled densely along its evolution variable, typically time, which the present single-plane measurements do not provide. More recently, physics-informed learning approaches,
such as Physics-Informed Neural Networks (PINNs) \cite{karniadakis2021physics,Ryu2024,DeRyck2024,Antonelo2024, zhang2025physics}, have attracted
considerable attention by combining experimental data with physical
constraints. Nevertheless, standard PINN formulations typically assume that the
governing partial differential equation (PDE) is known \emph{a priori} and employ neural networks to
approximate its solution or estimate unknown parameters. As a result,
they do not directly yield an explicit analytical input-output operator that
can be readily exploited for computationally efficient real-time optimal
control.

For transport-dominated DIDO systems, an intermediate route is therefore desirable: a model whose structure is selected from data among physically motivated transport equations, whose parameters are few and interpretable, and whose input--output operator is available in closed form for constrained online optimization.

Fan Array Wind Generators (FAWGs) constitute a representative platform
of DIDO systems for generating controllable turbulent wind fields. 
FAWGs are particularly suited to reproducing spatially and temporally varying wind conditions relevant to low-altitude flight, complementing conventional wind-tunnel studies of aerodynamics at higher flight regimes~\cite{kou2021data}. A typical FAWG consists of a large number of independently actuated fans whose neighboring jets interact aerodynamically \cite{Farell1999,Tieleman2003,Makita1991,Mydlarski1996,Noca2019,Azzam2023,cao2026turbulence}, producing strongly coupled spatial dynamics.
The kernel-based approach in Ref.~\onlinecite{shaqarin2025kernel} describes the relation between the fans' input and the measured velocity field, with an inverse model used for flow control. Although this approach showed good experimental performance, the model was mainly data-driven and did not explicitly describe the spatial evolution of the flow.

\begin{figure*}[t]
\centering
\includegraphics[width=\textwidth]{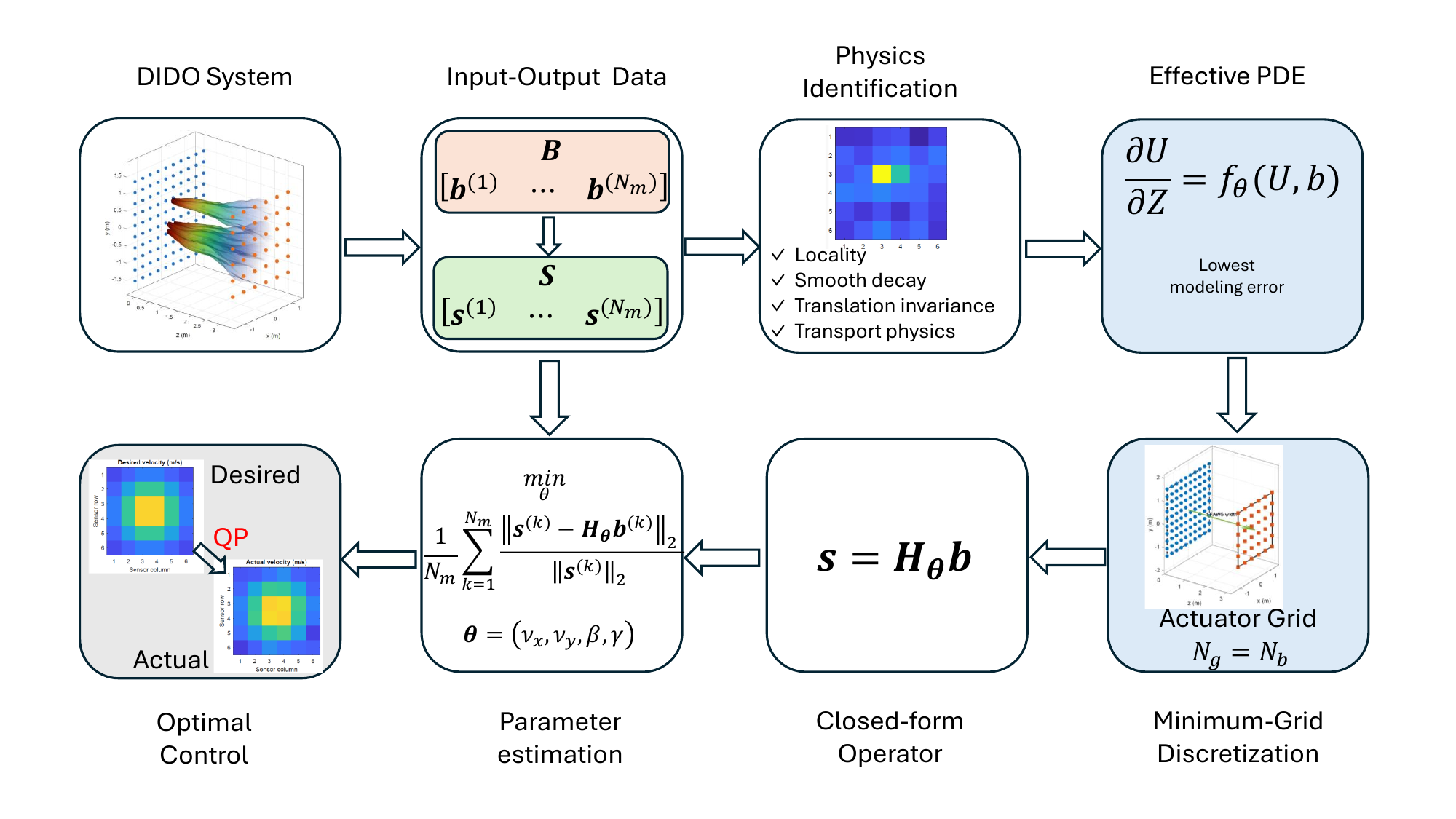}
\caption{
Overview of the proposed physics-based framework for modeling and optimal control of distributed-input distributed-output (DIDO) systems. Experimental input-output data are analyzed to identify the underlying transport physics from their spatial characteristics, including locality\rvdel{,} and approximate translation invariance. The identified effective PDE is discretized directly on the actuator grid, yielding a minimum-grid representation from which a closed-form input-output operator is derived. The resulting analytical model is subsequently used for parameter estimation and constrained optimal control of distributed fields.
}
\label{fig:framework}

\end{figure*}

In this work, we develop a physics-based framework for modeling and
optimal control of DIDO systems
from experimental data. The framework identifies an effective PDE describing the observed dynamics, discretizes it directly on
the actuator grid to obtain a minimum-grid representation, derives a closed-form input-output operator, estimates the governing parameters,
and enables constrained closed-loop optimal control, as summarized in Fig.~\ref{fig:framework}. Its effectiveness is
demonstrated experimentally on the inner $40\times40$ section of the world's highest-resolution FAWG. Compared with the recent data-driven FAWG model of Ref.~\onlinecite{stefan2026data}, the framework provides a two-dimensional physics-based description of the spatial interaction between the actuators and the measured flow field and covers all 1600 fans of this section.

The contributions are twofold. First, we show that the input--output map of a transport-dominated DIDO system can be identified by selecting among candidate transport PDEs discretized on the actuator grid, using only steady input--output data at a single sensor plane. Second, we show experimentally that the resulting four-parameter physics-based operator suffices for the iterative constrained control of a distributed velocity field using 1600 actuators.

The manuscript is organized as follows. Section~\ref{sec:methods} describes the experimental setup and the methodology for modeling and control. Section~\ref{sec:res} presents the identified transport model, its uncertainty, and its predictive and closed-loop performance. Section~\ref{sec:derivation} derives the structure of the identified transport equation from first principles. Section~\ref{sec:concl} summarizes the main findings and outlines future directions.

\section{Methods}\label{sec:methods}
% \rvdel{
% The proposed framework consists of four
% main stages. An effective transport PDE is first identified directly from experimental input-output data. The identified model is then solved
% in closed form to obtain an input--output operator, whose parameters are
% estimated from experiments and subsequently employed for constrained
% optimal control.}

\subsection{Experimental setup and dataset}

\begin{figure*}[t]
    \centering
    \includegraphics[width=1\textwidth]{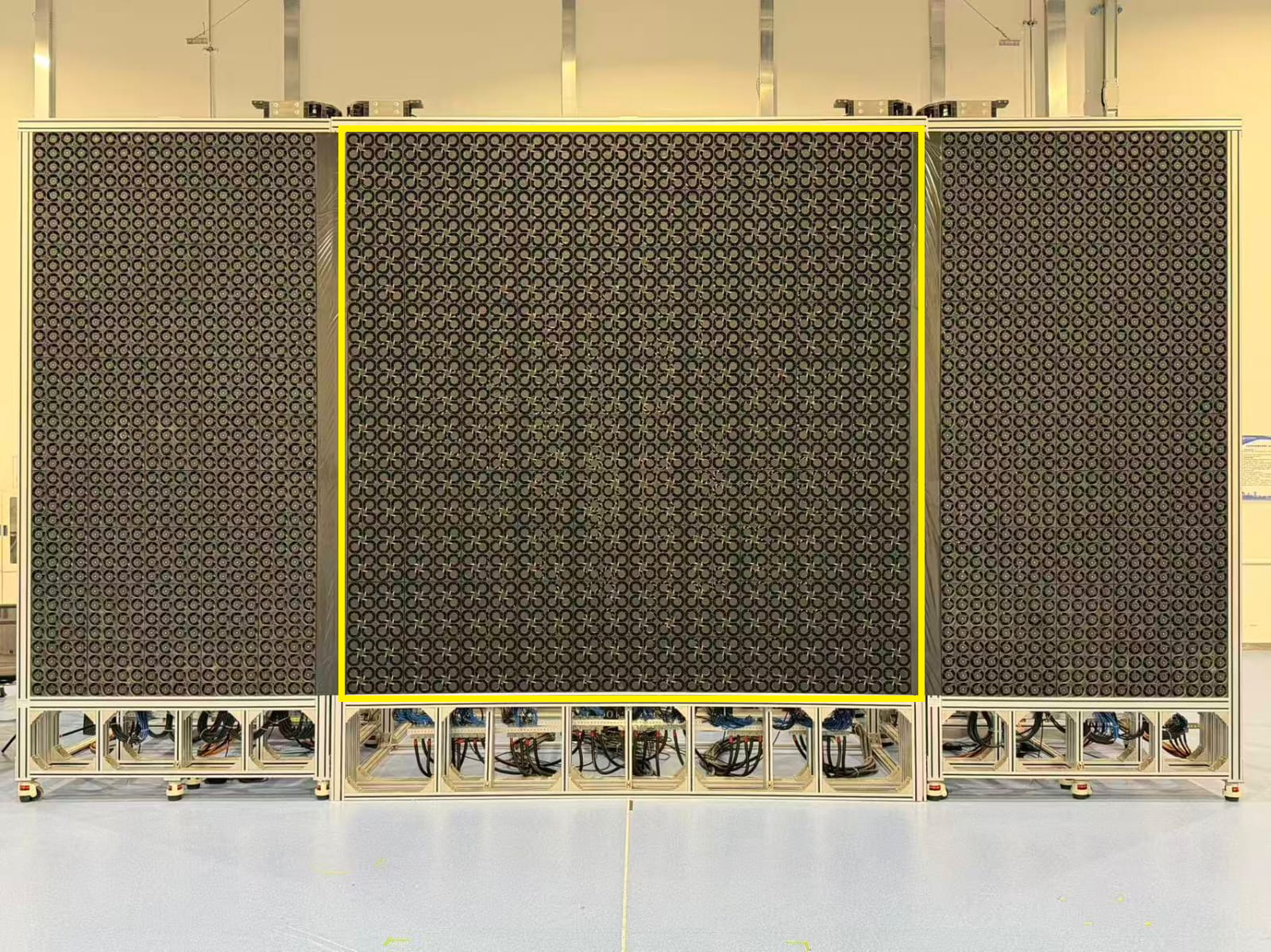}
    \caption{
    Experimental Fan Array Wind Generator (FAWG), consisting of
    $40\times80$ independently controlled fans. The inner $40\times40$
    fan array, indicated by the yellow square, is used in the present
    study.
    }
    \label{fig:FAWG}
\end{figure*}

The experiments were conducted at the Aerial City Lab, located within
the Shenzhen UAV Public Testing Center of Shenzhen Technology University.
The facility and the original configuration of the FAWG have been described in detail previously
\cite{liu2025aerodynamic,shaqarin2025kernel,cao2026turbulence}. 
The FAWG is a custom-built system (Dongteng Technology, China) consisting
of $40\times80$ independently controlled fan units, as shown in
Fig.~\ref{fig:FAWG}. In the present study, only the inner $40\times40$
section, indicated by the yellow square in Fig.~\ref{fig:FAWG}, is used.
This section comprises 1600 DC fans, each with a cross-section of
$80\times80~\mathrm{mm^2}$, distributed over an active area of
approximately $3.25\times3.25~\mathrm{m^2}$.

% providing 1600 spatially distributed flow
% actuators over an approximately $3.25\times3.25~\mathrm{m^2}$ square
% area. The array is mounted approximately $0.6~\mathrm{m}$ above the
% laboratory floor.

The fans are organized into modular $4\times4$ units, with each module connected to a dedicated 16-channel driver board that enables independent
pulse-width-modulation (PWM) control of every fan. The dual-rotor,
ball-bearing fans operate at 48~V DC with a rated rotational speed of
$14\,300$~RPM.
For the inner $40\times40$ section considered in this study, fan commands are transmitted from a host PC to two control boxes through
an RS-485 communication architecture, with each control box supervising
800 fans.  The control boxes forward the commanded PWM setpoints to the
corresponding driver boards using a Modbus-based RS-485 protocol.
Electrical power is supplied by five industrial AC--DC power supplies,
each rated at 48~V/1100~A (52.8~kW), providing a total installed capacity
of 264~kW.
The current configuration provides an operating velocity range
of approximately $2$--$18~\mathrm{m\,s^{-1}}$ under the conditions
considered in this study. The bottom edge of the fan array is located
approximately $0.6$~m above the laboratory floor. The array discharges into an enclosed hall, whose recirculating room flow has been characterized in Ref.~\onlinecite{wang2024coarse}. 

The streamwise velocity field is measured using the existing
$8\times8$ Pitot Tube Array (PTA). Only the inner $6\times6$ probes are used in the present modeling and control experiments, providing 36 spatially distributed velocity measurements while reducing the influence
of the outer shear layer and edge effects. The probes of the inner lattice are spaced by
$\Delta x=\Delta y=0.4647~\mathrm{m}$. The measurement plane is located $3.25~\mathrm{m}$ downstream of the
fan-array plane, corresponding to one width of the active section, or $Z=1$, where
$Z=z/W$ and $W=3.25~\mathrm{m}$. The transport models below are written in terms of $Z$. The same measurement setup was also used in Refs.~\onlinecite{liu2025aerodynamic,shaqarin2025kernel}. The dataset comprises 380 steady input--output experiments, including
20 isolated-actuator cases, 20 uniform-command cases, 20 directional
gradient cases, 240 fully populated random-command cases, and 80
additional structured spatial-pattern cases with different active
regions and command distributions. The actuator commands span
0--100\% PWM. For each experiment, a 2~s settling time was allowed
before measurement, after which the velocity measurements were averaged
over 10~s.

\begin{figure*}[t]
    \centering
    \includegraphics[width=0.75\textwidth]{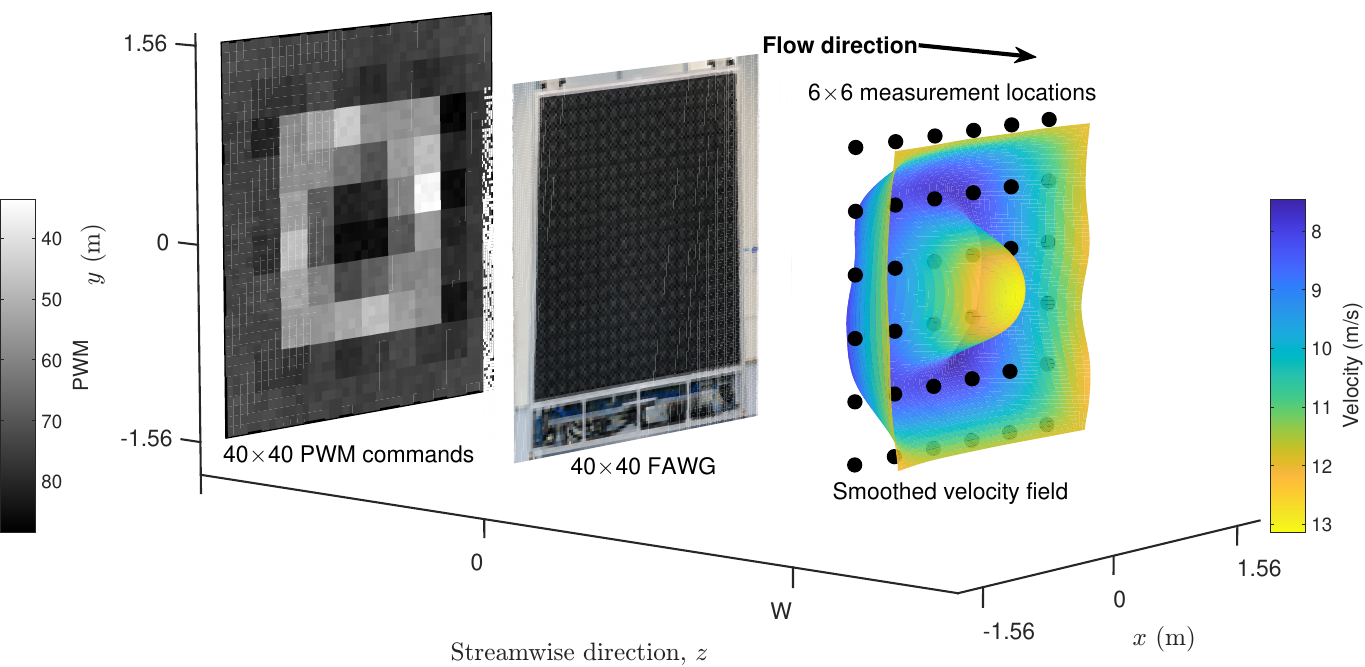}
    \caption{
    Experimental configuration and distributed input--output representation of the FAWG. The $40\times40$ fan PWM distribution is applied to the experimental FAWG. The resulting streamwise velocity field is measured by the inner $6\times6$ probes of the Pitot Tube Array (PTA), whose sensor locations are indicated by the black markers. The smoothed surface is an interpolation of the 36 measurements for visualization only.
    }
    \label{fig:FAWG_setup}
\end{figure*}

%\subsection{Distributed Input-Output Representation}

\subsection{Distributed input--output representation}
Consider a DIDO system, as illustrated in Fig.~\ref{fig:FAWG_setup}, consisting of $N_b$ spatially distributed
actuators and $N_s$ measurement locations. For a single experiment, the
actuator commands are collected in the vector
\begin{equation}
\mathbf{b}=
\begin{bmatrix}
b_1 & b_2 & \cdots & b_{N_b}
\end{bmatrix}^{T}.
\end{equation}
The measured physical field is represented by
\begin{equation}
\mathbf{s}=
\begin{bmatrix}
s_1 & s_2 & \cdots & s_{N_s}
\end{bmatrix}^{T}.
\end{equation}

For the FAWG considered in this work, the actuator vector corresponds to the individual fan PWM commands, whereas the output vector contains the measured velocity values at the sensing locations. The objective is to determine the actuator distribution required to generate a prescribed spatial field while accounting for the coupling between neighboring actuators.

Rather than assuming an \emph{a priori} governing equation, the
identification procedure begins by collecting experimental input-output
measurements from the distributed system. Performing $N_m$ experiments yields the input and output data matrices
\begin{equation}
\mathbf{B}=
\begin{bmatrix}
| & | & & |\\
\mathbf{b}^{(1)} & \mathbf{b}^{(2)} & \cdots & \mathbf{b}^{(N_m)}\\
| & | & & |
\end{bmatrix}, \,
\mathbf{S}=
\begin{bmatrix}
| & | & & |\\
\mathbf{s}^{(1)} & \mathbf{s}^{(2)} & \cdots & \mathbf{s}^{(N_m)}\\
| & | & & |
\end{bmatrix}.
\end{equation}
The empirical input--output operator is computed as

\begin{equation}
\label{eq:pinv}
\mathbf{G}=\mathbf{S}\mathbf{B}^{\dagger}\rvmdel{.}\rvadd{,}
\end{equation}
where $(\cdot)^{\dagger}$ denotes the Moore-Penrose pseudoinverse. 

\subsection{Candidate transport models}\label{sec:candidates}

Based on the observed spatial characteristics of the empirical
input--output operator (Sec.~\ref{sec:ident}), four candidate PDEs representing different
transport mechanisms were investigated. The models were selected to
examine the contributions of streamwise evolution, transverse advection,
and transverse diffusion to the spatial input--output behavior.

The first candidate assumes a steady-state transverse advection process,
\begin{equation}
v_x\frac{\partial U}{\partial x}
+
v_y\frac{\partial U}{\partial y}
=
\gamma b,
\end{equation}
where $v_x$ and $v_y$ represent effective transverse advection
coefficients, $\gamma$ is an effective actuator gain, and $b(x,y)$ is the actuator-command field. This model describes spatial transport across the
measurement plane without explicitly accounting for streamwise evolution
or transverse diffusion. Transverse advection would represent a systematic lateral drift of the jets, for instance due to residual swirl, to the recirculating room flow or to an asymmetric entrainment at the edges of the array.

The second candidate introduces streamwise evolution while retaining
transverse advection,
\begin{equation}
\frac{\partial U}{\partial Z}
+
v_x\frac{\partial U}{\partial x}
+
v_y\frac{\partial U}{\partial y}
=
\gamma b.
\end{equation}
Here, the dimensionless streamwise coordinate $Z$ acts as the evolution variable,
allowing the velocity field to develop between the actuator and
measurement planes.

The third candidate considers steady-state transverse
advection--diffusion,
\begin{equation}
v_x\frac{\partial U}{\partial x}
+
v_y\frac{\partial U}{\partial y}
=
\nu_x\frac{\partial^2U}{\partial x^2}
+
\nu_y\frac{\partial^2U}{\partial y^2}
+
\gamma b,
\end{equation}
where $\nu_x$ and $\nu_y$ represent effective transverse diffusion
coefficients. This formulation accounts for lateral spreading of the
actuator influence but does not explicitly describe its streamwise
evolution. In the two steady candidates, multiplying all coefficients by the same factor leaves the solution unchanged, so that only their ratios to $\gamma$ can be identified: these models have two and four independent identifiable parameters, respectively.

Finally, the anisotropic parabolic transport model combines streamwise
evolution and transverse diffusion,
\begin{equation}
\label{eq:anis_trans}
\frac{\partial U}{\partial Z}
=
\nu_x\frac{\partial^2U}{\partial x^2}
+
\nu_y\frac{\partial^2U}{\partial y^2}
-
\beta U
+
\gamma b,
\end{equation}
where $\beta$ represents an effective linear decay during streamwise
propagation. This model describes streamwise development together with
anisotropic transverse spreading and effective decay. Since $Z$ is dimensionless, $\nu_x$ and $\nu_y$ have units of m$^2$, $\beta$ is dimensionless, and $\gamma$ has units of m\,s$^{-1}$ per \% PWM.

Each candidate model was discretized on the same actuator grid, and its
unknown parameters were identified from the experimental input--output
dataset. First-order spatial derivatives were discretized using upwind or central finite
differences, as appropriate to each candidate model. The models were then compared using the mean normalized modeling
error defined in Eq.~\eqref{eq:parameter}. 
%\sdtodo{State the discretization of the first derivatives and the edge conditions of each candidate.}

%Each model was discretized on the same actuator grid, its unknown parameters were identified directly from the experimental input-output data, and the resulting predictions were compared with the measured velocity fields. The comparison was performed using the mean normalized prediction error computed over the entire experimental dataset.
\subsection{Minimum-grid PDE discretization}
To obtain a computationally efficient physics-based model, the governing
PDE is discretized directly on the actuator grid rather than on a finer
computational mesh. This minimum-grid discretization reduces the PDE
state dimension to the number of actuators ($N_g=N_b$) while preserving
the underlying spatial coupling required for prediction and control.
%The governing PDE is discretized directly on the actuator grid, yielding a minimum-state representation with $N_g=N_b$ while preserving the spatial coupling required for prediction and control.

Using second-order central finite differences, Eq.~\eqref{eq:anis_trans} becomes
\begin{equation}
\frac{d\mathbf{U}}{dZ}
=
\mathbf{A}_{\boldsymbol{\theta}}\mathbf{U}
+\gamma\mathbf{b},
\qquad
\mathbf{A}_{\boldsymbol{\theta}}
=
\nu_x\mathbf{L}_x
+\nu_y\mathbf{L}_y
-\beta\mathbf{I},
\label{eq:discreteODE}
\end{equation}
where $\mathbf{L}_x$ and $\mathbf{L}_y$ are the finite-difference
Laplacian operators in the spanwise and vertical directions,
respectively, $\mathbf{T}_x$ and $\mathbf{T}_y$ are the corresponding
one-dimensional second-derivative operators, and $\mathbf{I}$ is the
identity matrix.The grid spacing is $\Delta x=\Delta y=0.08$~m, with a zero-value exterior closure at the array edges. 
%\rvtodo{Confirm (i) the grid spacing ($W/N_x=0.081$~m; the fan positions in Fig.~3 suggest 0.080~m), (ii) the edge condition, and (iii) that $\mathbf{H}_{\boldsymbol{\theta}}$ is evaluated at $Z=1$; together they fix the units and values of $\nu_x$ and $\nu_y$ and the estimate in Sec.~IV.}

% \begin{equation}
% T_x=
% \frac{1}{\Delta x^2}
% \begin{bmatrix}
% -2 & 1 &        & 0\\
%  1 & -2 & \ddots &  \\
%    & \ddots & \ddots & 1\\
%  0 &        & 1 & -2
% \end{bmatrix},
% \qquad
% T_y=
% \frac{1}{\Delta y^2}
% \begin{bmatrix}
% -2 & 1 &        & 0\\
%  1 & -2 & \ddots &  \\
%    & \ddots & \ddots & 1\\
%  0 &        & 1 & -2
% \end{bmatrix},
% \end{equation}
% where $\Delta x$ and $\Delta y$ are the actuator-grid spacings in the
% spanwise and vertical directions, respectively.
Using the Kronecker product, the two-dimensional discrete
second-derivative operators are constructed as
\begin{equation}
\mathbf{L}_x
=
\mathbf{I}_{N_y}\otimes\mathbf{T}_x,
\qquad
\mathbf{L}_y
=
\mathbf{T}_y\otimes\mathbf{I}_{N_x},
\end{equation}
where $\mathbf{I}_{N_x}$ and $\mathbf{I}_{N_y}$ denote identity
matrices of dimensions $N_x$ and $N_y$, respectively. For the present
FAWG, $N_x=N_y=40$, giving $N_g=N_xN_y=1600$ states.

% Assuming constant actuation along the propagation direction, the closed-form solution of the discretized model at the measurement plane is obtained using the variation-of-constants formula,
% \begin{equation}
% U(L)
% =
% \gamma
% \int_{0}^{L}
% e^{A(L-\zeta)}\,d\zeta\, b.
% \end{equation}
% When $A$ is nonsingular, this expression reduces to
% \begin{equation}
% U(L)
% =
% \gamma
% A^{-1}
% \left(
% e^{AL}-I
% \right)b.
% \end{equation}
Assuming that the actuator-command vector $\mathbf{b}$ is constant along
the streamwise evolution coordinate and taking $\mathbf{U}(0)=\mathbf{0}$,
the solution of Eq.~\eqref{eq:discreteODE} at the measurement plane
follows from the variation-of-constants formula,
\begin{equation}
\mathbf{U}(Z)
=
\gamma
\int_{0}^{Z}
e^{\mathbf{A}_{\boldsymbol{\theta}}(Z-\zeta)}\,d\zeta\,\mathbf{b}
=
\gamma
\mathbf{A}_{\boldsymbol{\theta}}^{-1}
\left(
e^{\mathbf{A}_{\boldsymbol{\theta}}Z}-\mathbf{I}
\right)\mathbf{b},
\label{eq:UZ}
\end{equation}
where the second equality holds for nonsingular
$\mathbf{A}_{\boldsymbol{\theta}}$. In general, the integral equals $Z\,\varphi_1(\mathbf{A}_{\boldsymbol{\theta}}Z)$ with $\varphi_1(\mathbf{M})=\sum_{k\ge0}\mathbf{M}^k/(k+1)!$, which remains well defined when $\mathbf{A}_{\boldsymbol{\theta}}$ is singular. The distributed measurements at
the measurement plane ($Z=1$ in the present setup) are obtained through the interpolation operator
\begin{equation}
\mathbf{s}=\mathbf{C}\mathbf{U}(Z),
\end{equation}
where
$\mathbf{C}\in\mathbb{R}^{N_s\times N_g}$ maps the actuator-grid solution
to the sensor locations. Each row of $\mathbf{C}$ contains normalized
inverse-distance interpolation weights associated with the four nearest
grid nodes, such that the measurement at sensor $i$ is approximated by
a weighted average of the neighboring PDE states,
\begin{equation}
s_i=\sum_{j=1}^{N_g}C_{ij}U_j,
\qquad
\sum_{j=1}^{N_g}C_{ij}=1.
\end{equation}
% where the interpolation weights satisfy
% \begin{equation}
% \sum_{j=1}^{N_b}C_{ij}=1,
% \end{equation}
%thereby ensuring consistency between the discretized PDE solution and the sensor measurements. 
Combining Eq.~\eqref{eq:UZ} with the interpolation operator yields the closed-form input--output relation
\begin{equation}
\mathbf{s}
=
\mathbf{H}_{\boldsymbol{\theta}}\mathbf{b},
\end{equation}
where
\begin{equation}
\mathbf{H}_{\boldsymbol{\theta}}
=
\gamma\mathbf{C}
\int_{0}^{Z}
e^{\mathbf{A}_{\boldsymbol{\theta}}(Z-\zeta)}\,d\zeta
=
\gamma\mathbf{C}\mathbf{A}_{\boldsymbol{\theta}}^{-1}
\left(
e^{\mathbf{A}_{\boldsymbol{\theta}}Z}
-\mathbf{I}
\right),
\label{eq:H}
\end{equation}
or, in general, $\mathbf{H}_{\boldsymbol{\theta}}=\gamma Z\,\mathbf{C}\,\varphi_1(\mathbf{A}_{\boldsymbol{\theta}}Z)$.
% \begin{equation}
% H
% =
% \gamma C
% \int_{0}^{L}
% e^{A(L-\zeta)}\,d\zeta\,.
% \end{equation}
% When $A$ is nonsingular, this expression reduces to
% \begin{equation}
% H=
% \gamma
% CA^{-1}
% \left(
% e^{AL}-I
% \right).
% \label{eq:H}
% \end{equation}
The resulting operator provides a direct mapping between distributed actuator commands and sensor measurements without requiring numerical solution of the governing PDE during online operation.

\subsection{Parameter estimation}\label{sec:paramest}

The unknown physical parameters of Eq.~\eqref{eq:anis_trans} are
collected in the parameter vector
\begin{equation}
\boldsymbol{\theta}
=
\begin{bmatrix}
\nu_x &
\nu_y &
\beta &
\gamma
\end{bmatrix}^{T}.
\end{equation}
They are estimated by minimizing the discrepancy between the
experimentally measured outputs and those predicted by the analytical
model. The parameter estimation problem is formulated as
\begin{equation}
\min_{\boldsymbol{\theta}}\;
\frac{1}{N_m}
\sum_{k=1}^{N_m}
\frac{
\left\|
\mathbf{s}^{(k)}
-
\mathbf{H}_{\boldsymbol{\theta}}\mathbf{b}^{(k)}
\right\|_2
}{
\left\|
\mathbf{s}^{(k)}
\right\|_2
}.
\label{eq:parameter}
\end{equation}
The parameters were identified using the Nelder--Mead algorithm
(\texttt{fminsearch}), initialized at
$[\nu_x,\nu_y,\beta,\gamma]^T=[0.1,0.1,0.1,0.1]^T$; positivity was
enforced through a logarithmic parameterization, without explicit bounds. 
Because each experiment is normalized by its own output norm, Eq.~\eqref{eq:parameter} gives equal weight to experiments with weak and strong responses; the sparse, isolated commands, whose measured velocities are small, therefore may have a non-negligible influence on the fit and on the reported errors.

\subsection{Constrained optimal control}
Once the analytical operator has been identified, the generation of a desired distributed field is formulated as a constrained optimization problem. Equation~\eqref{eq:QP} below is its one-shot closed-loop form; the experiments use the incremental, closed-loop form of Sec.~\ref{sec:climpl}.

Given the desired output $\mathbf{s}_d$, the actuator commands are
determined by solving
\begin{equation}
\begin{aligned}
\min_{\mathbf{b}}\quad
&
\frac{1}{2}
\left\|
\mathbf{H}_{\boldsymbol{\theta}}\mathbf{b}
-\mathbf{s}_d
\right\|_2^2
+
\frac{\lambda}{2}
\|\mathbf{b}\|_2^2
\\
\text{subject to}\quad
&
\mathbf{0} \leq \mathbf{b} \leq b_{\max}\mathbf{1},
\end{aligned}
\label{eq:QP}
\end{equation}
where $\lambda>0$ regularizes the actuator effort. Since the Hessian
\begin{equation}
\mathbf{Q}
=
\mathbf{H}_{\boldsymbol{\theta}}^{T}
\mathbf{H}_{\boldsymbol{\theta}}
+\lambda\mathbf{I}
\end{equation}
is positive definite, the constrained optimization problem, a quadratic program (QP), is strictly
convex and admits a unique global solution. Convex QPs of this form are
computationally efficient and well suited for real-time control
implementation~\cite{galloway2015torque,shaqarin2024enhancing}. In the present implementation, the QP solution time was less than
$1~\mathrm{ms}$ on a standard PC.

\subsection{Closed-loop experimental implementation}\label{sec:climpl}
\begin{figure}[t]
\centering
\includegraphics[width=\columnwidth]{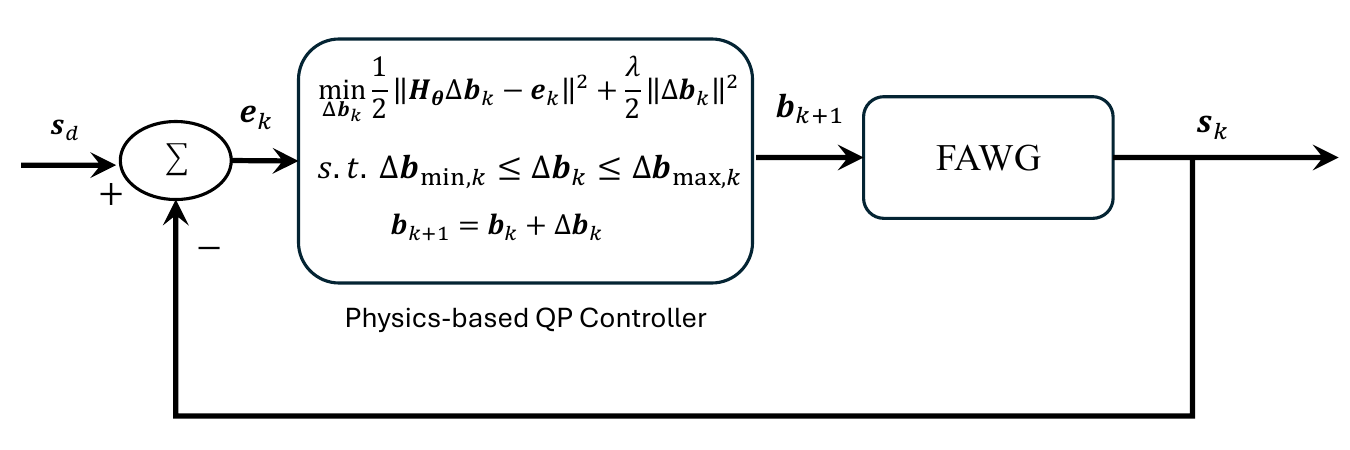}
\caption{
Closed-loop implementation of the proposed physics-based QP controller
using the analytical operator $\mathbf{H}_{\boldsymbol{\theta}}$ and
the measured tracking error.
}
\label{fig:controller}
\end{figure}

For the experimental closed-loop implementation, the model-based optimization is reformulated in incremental form (Fig.~\ref{fig:controller}). At iteration $k$, the tracking error is defined as
% \begin{equation}
% e_k=s_d-s_k.
% \end{equation}
% The incremental actuator command is obtained from
% \begin{equation}
% \begin{aligned}
% \min_{\Delta b_k}\quad
% &
% \frac{1}{2}
% \left\|
% H_\theta \Delta b_k-e_k
% \right\|_2^2
% +
% \frac{\lambda}{2}
% \left\|
% \Delta b_k
% \right\|_2^2
% \\
% \text{subject to}\quad
% &
% -b_k\leq \Delta b_k \leq b_{\max}-b_k .
% \end{aligned}
% \label{eq:incremental_QP}
% \end{equation}
% The actuator command is then updated according to
% \begin{equation}
% b_{k+1}=b_k+\Delta b_k.
% \end{equation}
\begin{equation}
\mathbf{e}_k
=
\mathbf{s}_d-\mathbf{s}_k.
\end{equation}

The incremental actuator command is obtained from
\begin{equation}
\begin{aligned}
\min_{\Delta\mathbf{b}_k}\quad
&
\frac{1}{2}
\left\|
\mathbf{H}_{\boldsymbol{\theta}}\Delta\mathbf{b}_k
-\mathbf{e}_k
\right\|_2^2
+
\frac{\lambda}{2}
\left\|
\Delta\mathbf{b}_k
\right\|_2^2
\\
\text{subject to}\quad
&
-\mathbf{b}_k
\leq
\Delta\mathbf{b}_k
\leq
b_{\max}\mathbf{1}-\mathbf{b}_k .
\end{aligned}
\label{eq:incremental_QP}
\end{equation}

The actuator command is then updated according to
\begin{equation}
\mathbf{b}_{k+1}
=
\mathbf{b}_k+\Delta\mathbf{b}_k.
\end{equation}
The controller is initialized with a uniform fan-command vector
$\mathbf{b}_0=50\%\mathbf{1}$ PWM. The incremental bounds in
Eq.~\eqref{eq:incremental_QP} ensure that
$\mathbf{0}\leq\mathbf{b}_{k+1}\leq b_{\max}\mathbf{1}$ at every
iteration.
The feedback iterations compensate for model error and experimental
disturbances. The QP is solved using MATLAB \texttt{quadprog} with the
\texttt{interior-point-convex} algorithm.

% =====================================================================
\section{Results}\label{sec:res}
% =====================================================================

\subsection{Identification of the effective transport model}\label{sec:ident}

\begin{figure*}[t]
    \centering
    \includegraphics[width=0.85\textwidth]
    {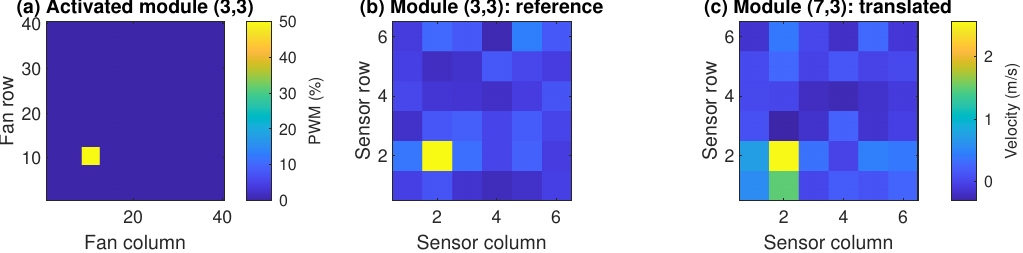}
    \caption{
    Spatial characteristics of the empirical input--output operator $\mathbf{G}$.
    (a) A localized $4\times4$ actuator module at position $(3,3)$ is
    excited at 50\% PWM.
    (b) The corresponding streamwise velocity footprint obtained from the empirical operator, demonstrating the localized spatial influence of the actuation.
    (c) Downstream footprint corresponding to module $(7,3)$ after translation to the reference location of module $(3,3)$.
    }
    \label{fig:empirical_locality}

\end{figure*}

The spatial characteristics of the empirical operator $\mathbf{G}$ in
Eq.~\eqref{eq:pinv} are examined to determine the underlying physical
behavior. As illustrated in Fig.~\ref{fig:empirical_locality}, $\mathbf{G}$
exhibits a localized actuator footprint
and approximate translation invariance. In particular,  localized
$4\times4$ actuator excitation produces a localized response at the
measurement plane, while translation of the actuator module results in
an approximately translated downstream footprint. The footprint in Fig.~\ref{fig:empirical_locality}(b) essentially reproduces the measured response to the corresponding module experiment. These spatial characteristics motivate representing the interaction
between neighboring actuators and the measured velocity field by a
local differential operator. 

Four candidate transport models were therefore compared using the mean normalized modeling error over the 380 experimental input--output sets.
As shown in Table~\ref{tab:model_comparison}, the steady-state advection model gives the largest error (59\%). Introducing streamwise evolution reduces the error to 33\%, while the steady-state advection--diffusion
model gives an error of 36\%. The lowest error, 26\%, is obtained with the anisotropic parabolic transport model, which combines streamwise evolution with transverse diffusion. This model, Eq.~\eqref{eq:anis_trans}, is selected as the effective transport model; its physical basis is discussed in Sec.~\ref{sec:derivation}. Streamwise evolution and transverse diffusion each reduce the error of the steady advection model, and together they give the lowest error.
\begin{table*}[t]
\centering
\caption{Comparison of the candidate physics-based models using 380
experimental input--output sets, including uniform, random, and isolated
input variations. 
%T\rvdel{Starting from the steady-state advection model, the inclusion of either streamwise evolution or transverse diffusion substantially improves the modeling accuracy. The lowest error is obtained when both mechanisms are included, supporting the anisotropic parabolic transport equation without lateral advection}
Mean error: Eq.~\eqref{eq:parameter}}.
\label{tab:model_comparison}

\renewcommand{\arraystretch}{1.6}
\setlength{\tabcolsep}{5pt}

\begin{tabular}{@{}l p{7.5cm} c c@{}}
\toprule
Model & Governing equation & Parameters & Mean error (\%) \\
\midrule

Steady-state advection
&
\(\displaystyle
v_x\frac{\partial U}{\partial x}
+
v_y\frac{\partial U}{\partial y}
=
\gamma b
\)
&
3
&
59
\\

Parabolic advection
&
\(\displaystyle
\frac{\partial U}{\partial Z}
+
v_x\frac{\partial U}{\partial x}
+
v_y\frac{\partial U}{\partial y}
=
\gamma b
\)
&
3
&
33
\\

Steady-state advection--diffusion
&
\(\displaystyle
v_x\frac{\partial U}{\partial x}
+
v_y\frac{\partial U}{\partial y}
=
\nu_x\frac{\partial^2 U}{\partial x^2}
+
\nu_y\frac{\partial^2 U}{\partial y^2}
+
\gamma b
\)
&
5
&
36
\\

Anisotropic parabolic transport
&
\(\displaystyle
\frac{\partial U}{\partial Z}
=
\nu_x\frac{\partial^2 U}{\partial x^2}
+
\nu_y\frac{\partial^2 U}{\partial y^2}
-
\beta U
+
\gamma b
\)
&
4
&
26
\\

\bottomrule
\end{tabular}
\end{table*}

\subsection{Parameter estimates and uncertainty}\label{sec:uncert}
Table~\ref{tab:parameter_uncertainty} lists the parameters of Eq.~\eqref{eq:anis_trans} identified from the complete data set.

\begin{figure}[t]
    \centering
    \includegraphics[width=\columnwidth]
    {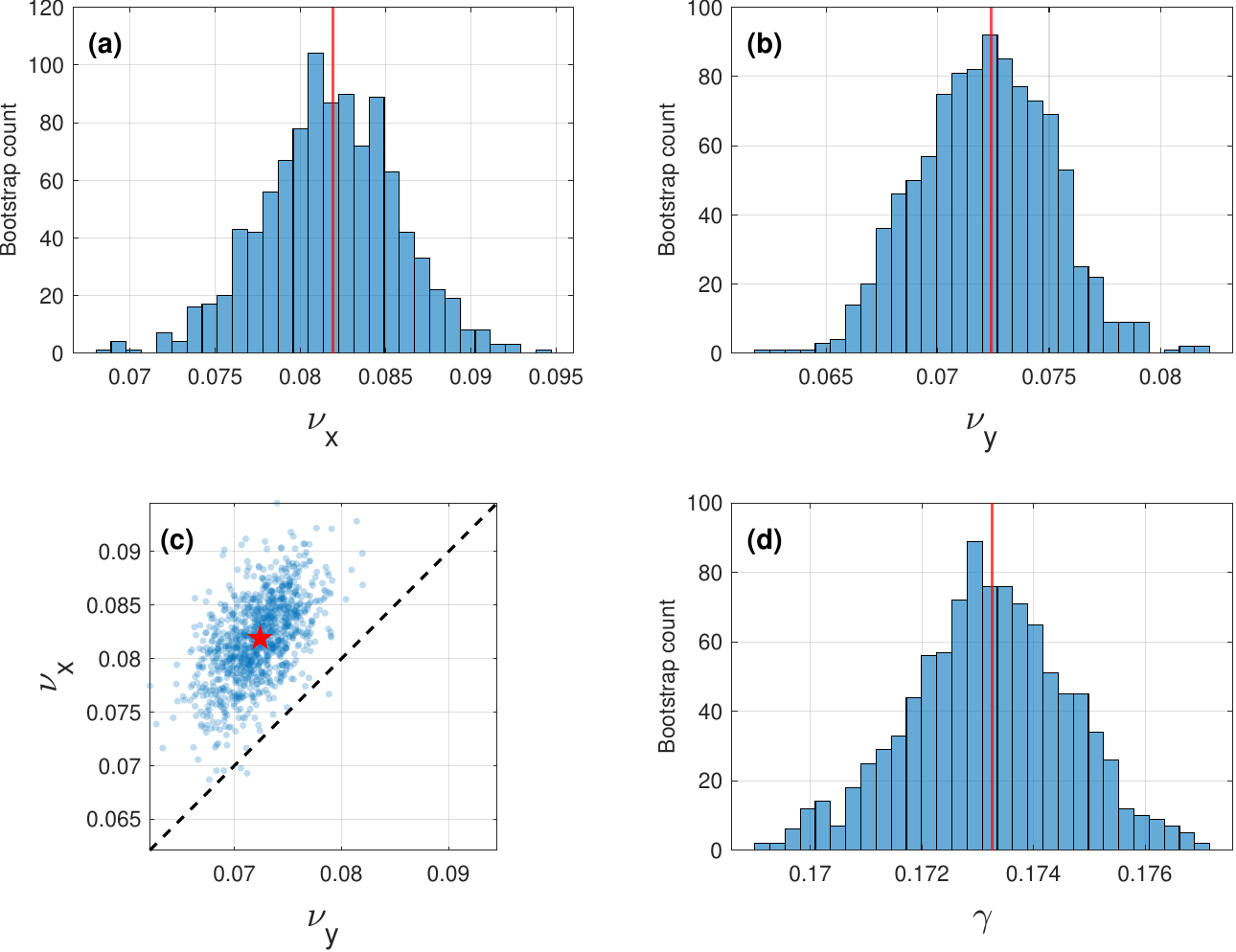}
    \caption{Bootstrap distributions of the identified model parameters
obtained from 1000 resampled experimental datasets:
(a) $\nu_x$, (b) $\nu_y$, (c) joint distribution of
$(\nu_x,\nu_y)$, and (d) $\gamma$.
The red vertical lines in (a), (b), and (d) indicate the parameter
estimates obtained from the complete experimental dataset. In (c),
the dashed line denotes $\nu_x=\nu_y$, and the red star indicates
the estimate obtained from the complete dataset.}
    \label{fig:bootstrap}
\end{figure}

\begin{table}[t]
\caption{Identified model parameters and 95\% confidence intervals
obtained from 1000 bootstrap realizations.}
\label{tab:parameter_uncertainty}
\centering
\begin{tabular}{lcc}
\hline\hline
Parameter & Estimate & 95\% bootstrap CI \\
\hline
$\nu_x$ (m$^2$\rvdel{/s})
& 0.0819
& $[0.0737,\;0.0891]$ \\

$\nu_y$ (m$^2$)
& 0.0724
& $[0.0666,\;0.0778]$ \\

$\beta$ ($-$)
& $1.68\times10^{-12}$
& $[2.01\times10^{-13},\;7.10\times10^{-12}]$ \\

$\gamma$ (m\,s$^{-1}$ per \% PWM\rvdel{ unit})
& 0.1733
& $[0.1702,\;0.1759]$ \\
\hline\hline
\end{tabular}
\end{table}
The uncertainty of the identified parameters was assessed using a
nonparametric bootstrap procedure. Complete command--measurement
experiments were resampled with replacement from the original dataset,
and the parameter-identification procedure was repeated for 1000
bootstrap realizations. The resulting parameter distributions are shown in
Fig.~\ref{fig:bootstrap}, while the corresponding 95\% confidence
intervals are summarized in Table~\ref{tab:parameter_uncertainty}.
The distributions of $\nu_x$, $\nu_y$, and $\gamma$ are centered close
to the estimates obtained from the complete dataset and have relatively
narrow confidence intervals. The bootstrap estimate of
$\nu_x/\nu_y$ is $1.131$, with a 95\% confidence interval of
$[1.046,1.232]$, which excludes unity and supports the identified
anisotropy.
In contrast, $\beta$ shows a larger
relative uncertainty because its nominal value is extremely close to zero.
Since $1-e^{-\beta Z}<10^{-11}$ for all values in the confidence interval, $\beta$ has no measurable effect and is best regarded as zero; setting $\beta=0$ leaves three parameters and the same mean error of 26\%. Note also that the bootstrap intervals quantify only the sampling variability of the fit; they do not account for the error of the model form, which dominates the 26\% residual.

\subsection{Forward-model assessment}\label{sec:forward}

% \rvdelblock{\rvdel{Discretization of the identified transport equation directly on the
% $40\times40$ actuator grid results in a 1600-state minimum-grid
% representation. The closed-form solution of the spatial evolution equation,
% combined with interpolation to the 36 measurement locations, yields the
% analytical input--output relation
% }\begin{equation}
% \mathbf{s}
% =
% \mathbf{H}_{\boldsymbol{\theta}}\mathbf{b},
% \end{equation}\rvdel{
% where
% $\mathbf{H}_{\boldsymbol{\theta}}\in\mathbb{R}^{36\times1600}$
% directly maps the fan commands to the measured velocity field.}}

% \rvdel{The structure of $\mathbf{H}_{\boldsymbol{\theta}}$ is determined by the
% identified transport model and depends on only four physical parameters,
% $\nu_x$, $\nu_y$, $\beta$, and $\gamma$, with the identified value of
% $\beta$ being effectively negligible over the considered propagation
% distance. Once these parameters are identified, evaluation of the
% distributed output requires only the matrix operation
% $\mathbf{H}_{\boldsymbol{\theta}}\mathbf{b}$, without numerical
% integration of the PDE.}

\begin{figure*}[t]
    \centering
    \includegraphics[width=\textwidth]
    {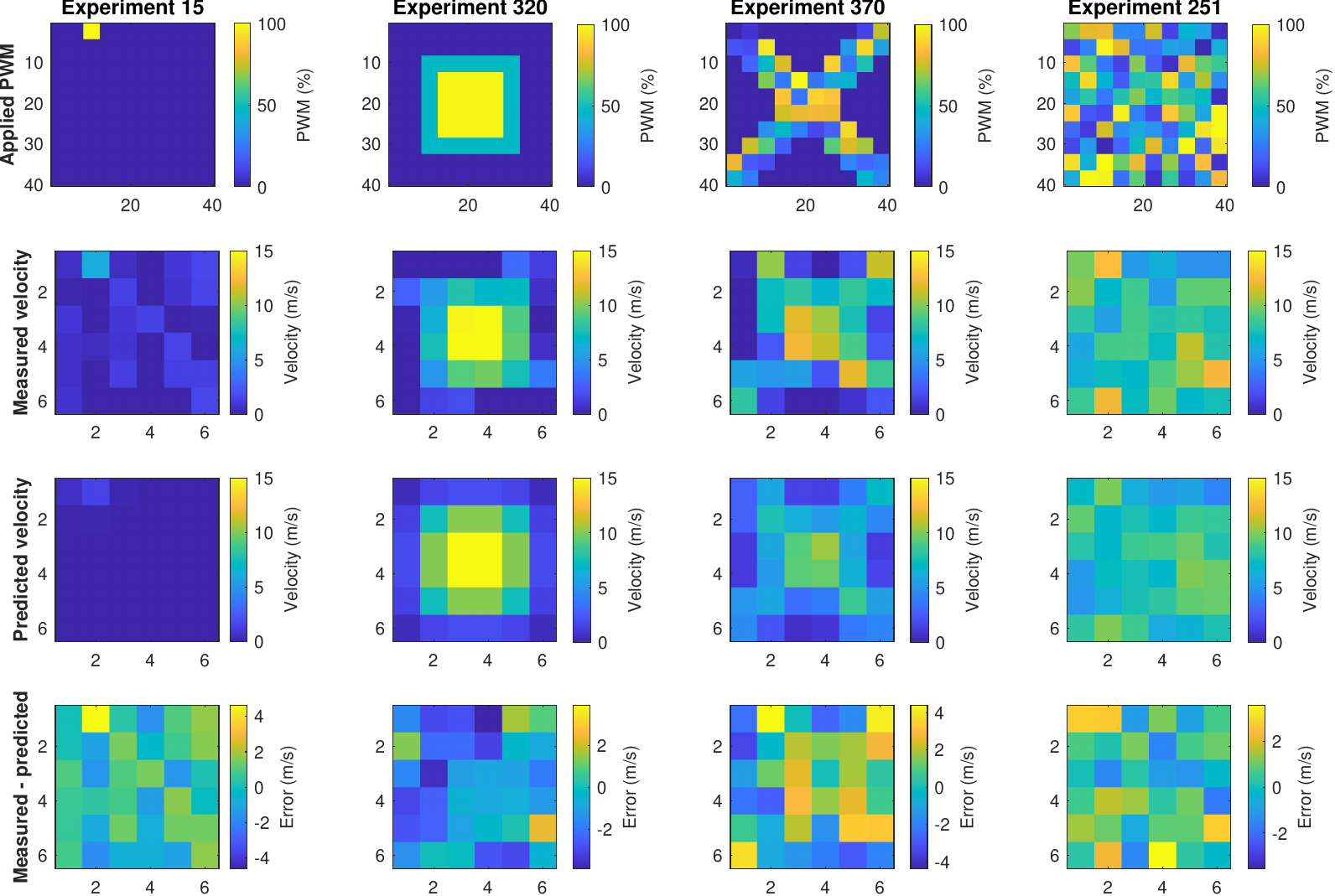}
    \caption{Physics-based model comparison for representative experimental
actuator distributions. The columns show the applied $40\times40$
PWM distribution, the experimentally measured velocity field, the
velocity field predicted by the physics-based operator
$\mathbf{H}_{\boldsymbol{\theta}}$, and the corresponding error
$\mathbf{s}_{\mathrm{exp}}-\mathbf{s}_{\mathrm{pred}}$.
    }
    \label{fig:forward_model}
\end{figure*}

To assess the proposed forward model, i.e., the identified analytical
operator $\mathbf{H}_{\boldsymbol{\theta}}$, four representative
experiments from the identification dataset are considered. These
include a concentrated point-like source (Exp.~15), centered localized
actuation (Exp.~320), an X-shaped actuation pattern (Exp.~370), and a
random actuation pattern (Exp.~251), as shown in
Fig.~\ref{fig:forward_model}.
The applied $40\times40$ fan commands are supplied
to $\mathbf{H}_{\boldsymbol{\theta}}$, and the predicted velocity fields are compared
with the corresponding measurements. 
The model captures the main spatial characteristics of the measured velocity fields, although quantitative differences remain. The corresponding mean absolute errors are $1.10$, $1.64$, $2.00$, and $1.15~\mathrm{m\,s^{-1}}$ for Exp.~15, Exp.~320, Exp.~370, and Exp.~251, respectively. For a consistent comparison, the mean absolute errors are normalized
by $10~\mathrm{m\,s^{-1}}$, corresponding to the midpoint of the
$2$--$18~\mathrm{m\,s^{-1}}$ operating velocity range of the FAWG.
The resulting normalized mean absolute errors range from $11.0\%$ to
$20.0\%$. These errors represent the forward-model error prior to closed-loop correction. They are not directly comparable with the 26\% of Table~\ref{tab:model_comparison}, which is the relative $\ell_2$ error of Eq.~\eqref{eq:parameter} averaged over all 380 experiments.

%\rvdel{The same operator can therefore be directly incorporated into the constrained optimal-control problem, where feedback is used to compensate for the model error.}

\subsection{Closed-loop generation of distributed velocity fields}\label{sec:closedloop}

%\subsection{Optimal Control Formulation}

% For the experimental closed-loop implementation, the QP controller
% in Eq.~\eqref{eq:QP} is applied incrementally, as illustrated in
% Fig.~\ref{fig:controller}. The controller is initialized with a uniform
% fan command $b_0=50\%$ PWM. At iteration $k$, the measured tracking error
% $e_k=s_d-s_k$ is used to determine an incremental fan command
% $\Delta b_k$, and the actuator command is updated as
% $b_{k+1}=b_k+\Delta b_k$. The incremental bounds are adjusted as
% $-b_k\leq\Delta b_k\leq b_{\max}-b_k$, ensuring
% $0\leq b_{k+1}\leq b_{\max}$ at every iteration.
% The regularization parameter $\lambda=10^{-8}$ was selected empirically
% to limit the actuator effort while maintaining tracking accuracy.
% The feedback iterations compensate for model error and experimental
% disturbances. The QP is solved using MATLAB \texttt{quadprog} with the
% \texttt{interior-point-convex} algorithm.

The framework is evaluated through
closed-loop experimental generation of representative distributed
velocity fields using the controller shown in
Fig.~\ref{fig:controller}. Three target distributions are considered, namely a
localized Gaussian profile, a two-dimensional sinusoidal distribution,
and a uniform velocity field, representing localized, spatially varying,
and homogeneous flow conditions, respectively. The Gaussian target has a center peak of approximately
$14~\mathrm{m\,s^{-1}}$ and a characteristic width of
$\sigma\approx0.65~\mathrm{m}$, the two-dimensional sinusoidal target
has an amplitude of $2.5~\mathrm{m\,s^{-1}}$ about a mean level of
$10.5~\mathrm{m\,s^{-1}}$ and a wavelength of $2.5~\mathrm{m}$ in
both spatial directions, and the uniform target is
$8~\mathrm{m\,s^{-1}}$.

\begin{figure*}[t]
\centering
\includegraphics[width=\textwidth]{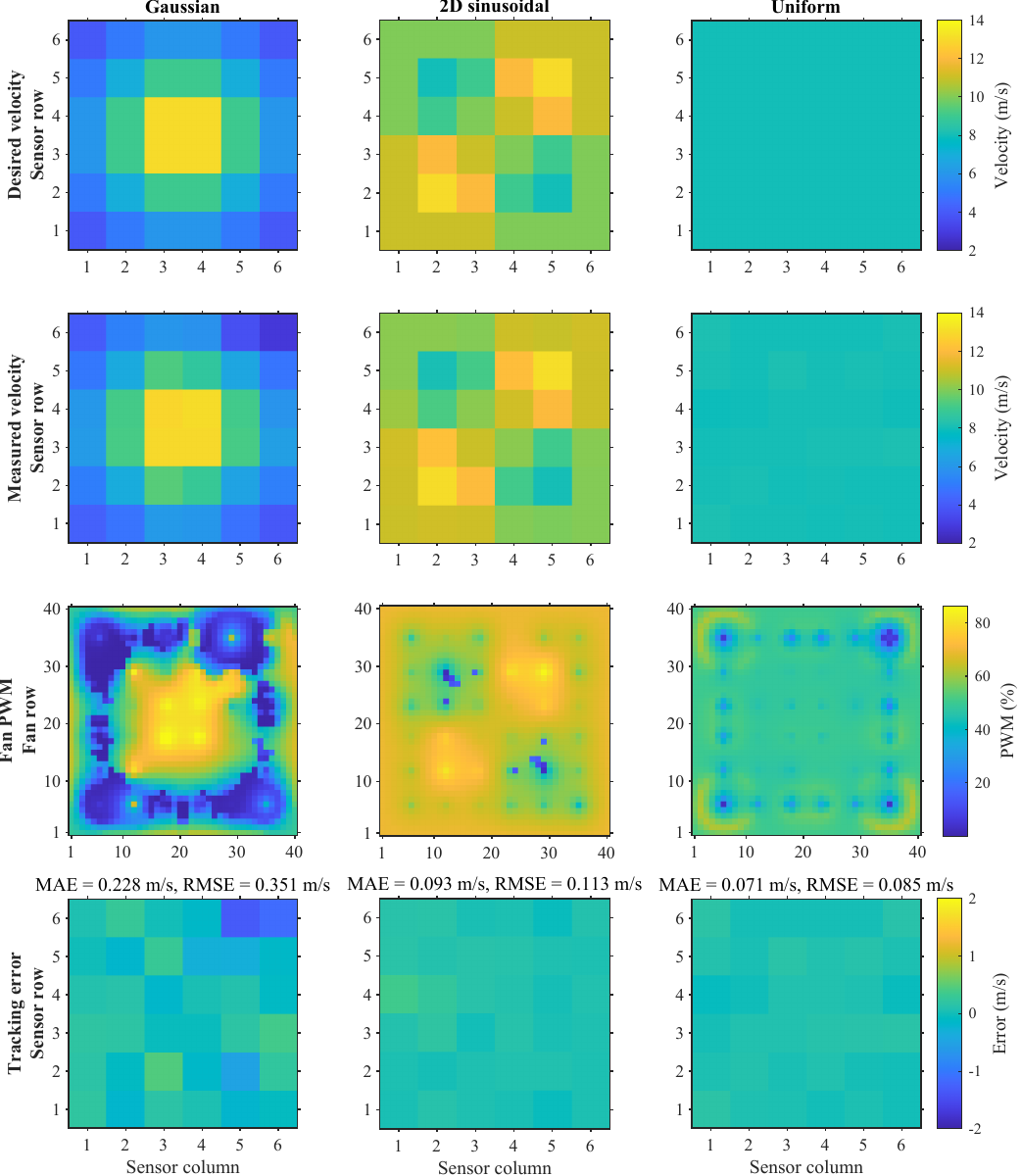}
\caption{
Closed-loop experimental generation of representative distributed velocity fields using the proposed framework. The three columns correspond to (a) a localized Gaussian profile, (b) a two-dimensional sinusoidal distribution, and (c) a uniform velocity field. For each case, the first row shows the desired velocity distribution, the second row the experimentally measured velocity field, the third row the optimized fan PWM distribution computed by the quadratic programming controller, and the fourth row the corresponding tracking error. 
}
\label{fig:closedloop}
\end{figure*}

Figure~\ref{fig:closedloop} shows that the controller
accurately reproduces all three target velocity distributions despite the strong aerodynamic coupling between neighboring
fans. The optimized PWM distributions exhibit nontrivial spatial
patterns, indicating that the controller exploits the
modeled inter-fan interactions to generate the desired velocity fields. Part of this structure, notably the imprint of the probe lattice in the uniform case, reflects the regularization, which penalizes the norm of the control increment $\Delta\mathbf{b}_k$ while minimizing the tracking error.

The corresponding mean absolute errors are $0.228$, $0.093$, and $0.071~\mathrm{m\,s^{-1}}$ for the localized Gaussian profile, the
two-dimensional sinusoidal distribution, and the uniform velocity field,
respectively; convergence was achieved after 10, 7, and
9 closed-loop iterations.
Using the same normalization adopted for the forward-model
assessment, the resulting normalized mean absolute errors range from
$0.7\%$ to $2.3\%$, substantially lower than the $11\%$--$20\%$ forward-model errors.
This reduction demonstrates the ability of the
closed-loop controller to compensate for forward-model mismatch while
generating the prescribed distributed velocity fields.

\section{First-principles derivation of the effective transport model}\label{sec:derivation}
% Although a Boussinesq-type transverse diffusion model provides the most
% natural physical candidate for the turbulent mixing generated by the
% fan array, it was not assumed a priori. Alternative transport
% mechanisms were also evaluated against the experimental data, as
% summarized in Table~\ref{tab:model_comparison}.
The PDE model \eqref{eq:anis_trans}, selected in Table~\ref{tab:model_comparison}, 
is well rooted in shear turbulence properties of the FAWG.
The structure of the best effective transport equation can be derived from
the Reynolds-averaged momentum equation of nearly unidirectional turbulent shear flow in the far-field approximation.
First, the Reynolds-averaged momentum equation is parabolized in the streamwise direction.
Second, the Reynolds stress is modeled by the Boussinesq eddy-viscosity hypothesis\cite{pope2000turbulent} 
with eddy viscosities $\nu_{t,x}$ and $\nu_{t,y}$ representing
momentum transfer in the two transverse directions.
Third, the eddy viscosities are assumed to only depend on the streamwise coordinate; for comparison with Eq.~\eqref{eq:anis_trans} they are further approximated by their average between the fan array and the measurement plane.
Fourth, the pressure gradient can be neglected in the far field.
Fifth, the fans are represented by a streamwise body force per unit mass, $F_b(x,y)$, proportional to the local command $b$. Sixth, the transverse mean velocities are neglected; consistently, the candidates relying on transverse advection perform worse than the anisotropic parabolic transport model in Table~\ref{tab:model_comparison}.
Under these approximations,
the streamwise momentum equation reads 
\begin{equation}
U\frac{\partial U}{\partial z}
=
\nu_{t,x}\frac{\partial^2U}{\partial x^2}
+
\nu_{t,y}\frac{\partial^2U}{\partial y^2}
+
F_b.
\label{eq:boussinesq_reduced}
\end{equation}
To obtain a linear streamwise-evolution model, the nonlinear convective
term may be linearized using an Oseen-type approximation, in which the
local streamwise convection velocity $U$ multiplying the streamwise
gradient is replaced by a characteristic convection velocity $U_c$.
Equation~\eqref{eq:boussinesq_reduced} then becomes
\begin{equation}
U_c\frac{\partial U}{\partial z}
=
\nu_{t,x}\frac{\partial^2U}{\partial x^2}
+
\nu_{t,y}\frac{\partial^2U}{\partial y^2}
+
F_b.
\label{eq:boussinesq_oseen}
\end{equation}
Dividing by $U_c$ gives
\begin{equation}
\frac{\partial U}{\partial z}
=
\frac{\nu_{t,x}}{U_c}
\frac{\partial^2U}{\partial x^2}
+
\frac{\nu_{t,y}}{U_c}
\frac{\partial^2U}{\partial y^2}
+
\frac{F_b}{U_c}.
\label{eq:boussinesq_oseen_normalized}
\end{equation}
Equation~\eqref{eq:boussinesq_oseen_normalized} has the same principal
transport structure as the experimentally identified effective model,
namely streamwise evolution, anisotropic transverse diffusion, and
distributed forcing. In particular, comparison with
Eq.~\eqref{eq:anis_trans} suggests the correspondence
\begin{equation}
\nu_x \sim W\frac{\nu_{t,x}}{U_c},
\qquad
\nu_y \sim W\frac{\nu_{t,y}}{U_c},
\qquad
\gamma b \sim W\frac{F_b}{U_c}.
\label{eq:physical_correspondence}
\end{equation}
The factor $W$ arises because Eqs.~\eqref{eq:boussinesq_reduced}--\eqref{eq:boussinesq_oseen_normalized} are written in the dimensional coordinate $z$, whereas the identified model uses $Z=z/W$. Notably, the identified decay coefficient is nearly zero
(Table~\ref{tab:parameter_uncertainty}). Thus, the experimental identification does
not require an appreciable additional linear decay term, and the
identified effective transport equation closely approaches the
Boussinesq--Oseen form in
Eq.~\eqref{eq:boussinesq_oseen_normalized}.

The Oseen linearization restricts Eq.~\eqref{eq:boussinesq_oseen} to moderate variations of the velocity around $U_c$. Without forcing, it conserves the transverse integral of $U$, whereas a free turbulent jet conserves its momentum flux and entrains ambient fluid through the transverse velocities neglected here; this is consistent with the larger errors for sparse commands in Sec.~\ref{sec:forward}.

The physical relevance of Eq.~\eqref{eq:boussinesq_oseen_normalized}
can be further corroborated by  a parameter estimate. 
According to Prandtl's mixing length theory, 
the eddy viscosity  $\nu_t=\nu_{t,x}=\nu_{t,y} = u_m \> l_m$
can be approximated by the eddy velocity $u_m$ 
and mixing length $l_m$.
We estimate the mixing length using the individual fan pitch, $l_m=0.08$~m, and the eddy velocity with a typical value of shear turbulence $u_m=0.2 U_c$. This gives $Wu_m l_m/U_c\approx0.052~\mathrm{m^2}$,
which is about 33\% smaller than the
average of both identified coefficients in Table~\ref{tab:parameter_uncertainty}.
Evidently, the parameter estimate can and should be refined.
Furthermore, the identified values correspond to $u_m/U_c=\nu_{x,y}/(W l_m)\approx0.28$--$0.32$, i.e., an eddy velocity of about 28--32\% of the convection velocity, , comparable to the
assumed 20\%. 
%This is plausible, since in mixing-length theory $u_m$ scales with the velocity difference between adjacent jets rather than with $U_c$ itself. 
The order-of-magnitude agreement between this first-principles estimate 
and the identified diffusion coefficients supports the physical interpretation of the identified PDE.

\section{Conclusions and outlook}\label{sec:concl}

We propose a physics-based framework for modeling and constrained optimal control of DIDO systems from experimental input--output data. The principal result of this work is that the high-dimensional input--output behavior of the 1600-fan array can be represented by a compact effective transport model containing only four identified parameters. Among the investigated candidate models, the combination of streamwise evolution and anisotropic transverse diffusion provides the closest agreement with the experimental data. The decay coefficient is effectively zero, indicating that transverse spreading dominates the modeled spatial interaction between the actuator and measurement plane. The remaining parameters have narrow bootstrap confidence intervals. The identified PDE is consistent with both the structure and the characteristic scaling of a parabolized Boussinesq--Oseen model (Sec.~\ref{sec:derivation}); alternative forcing and boundary-condition formulations may provide other effective representations of the measured input--output behavior.
%Furthermore, the present identification is based on measurements at a single downstream plane. Measurements at multiple downstream locations would allow the streamwise evolution predicted by different model formulations to be examined directly.

Unlike purely data-driven input--output models, such as the kernel model of Ref.~\onlinecite{shaqarin2025kernel}, the framework represents the spatial coupling through an explicit transport operator, which connects the identified spatial behavior with the model used for constrained optimal control. The closed-form operator $\mathbf{H}_{\boldsymbol{\theta}}$ allows the online controller to operate on the input--output mapping without numerically solving the PDE at each iteration, which is particularly important for high-dimensional actuation with 1600 independently controlled inputs. In closed loop, the constrained QP controller generates all three prescribed velocity fields in fewer than 10 iterations with normalized mean absolute errors of $0.7\%$--$2.3\%$, compared with 11--20\% for the forward model alone, while respecting the actuator limits explicitly.

Overall, the results demonstrate that a physics-based representation can retain the computational simplicity required for high-dimensional distributed control while providing a physically interpretable description of the spatial coupling between actuators and measurements. Although demonstrated here on a FAWG, the formulation can be extended to other distributed physical systems whose input--output behavior is dominated by spatial transport processes.

The presented work is a starting point for many future activities.
The measurement system used in this
work relies on Pitot tubes and mean velocity measurements, which limits its
ability to resolve transient flow dynamics. Future work will employ hot-wire anemometry with higher temporal resolution. In addition, a finer sensor
array will be deployed to support the complete $40\times80$ FAWG,
enabling the framework to be extended from the present 1600-fan
configuration to all 3200 independently controlled fans and allowing more
precise control of complex spatial flow fields.

%\rvdel{Preliminary work indicates that the PDE model is a good approximation for other FAWG using the same fans.Moreover, the selected effective transport equation describes also the flow field 1/2 width and 2 widths downstream.Both observations encourage a}
Measurements at several downstream planes and on other FAWGs will show which of the identified parameters carry over
and will support refined first-principles modeling
of the spatio-temporal dynamics of streamwise flow in a large volume. %\rvnote{Claims on other planes and other FAWGs removed: with $\beta\approx0$ the model gives $U(Z)=\gamma Z b$ for a uniform command (8.7, 17.3 and 34.7~m\,s$^{-1}$ at $Z=1/2$, 1 and 2 for $b=100\%$), so other planes cannot be described without refitting. Restore them only with a figure that uses the $Z=1$ parameters.}
%\sddel{Recent work by this team{\cite{cao2026turbulence}} indicates how the current framework may be generalized to predict mean flow and fluctuation levels simultaneously if physically realizable.}
Recent work on the same facility\cite{cao2026turbulence} suggests that the framework could be extended to mean velocity and turbulence levels simultaneously.
%\sddel{The PDE model opens a new avenue in modeling and control of FAWGs.}

% In summary, a physics-based framework has been proposed for modeling and
% optimal control of distributed-input distributed-output (DIDO) systems.
% Starting from experimental input-output data, an effective PDE
% is identified and discretized directly on the actuator grid, yielding a
% minimum-grid representation from which a closed-form analytical
% input-output operator is derived. The resulting model requires only four physically interpretable parameters and provides
% an efficient formulation for constrained closed-loop optimal control. The identified model indicates that the dominant transport mechanisms are streamwise evolution and anisotropic transverse diffusion, while the fitted dissipation coefficient is negligible over the considered propagation distance.
% Experimental results on a $40\times40$ fan-array wind generator
% demonstrate accurate generation of representative distributed velocity
% fields and demonstrate the effectiveness of the proposed physics-based modeling and control framework. Owing to its general
% formulation, the proposed framework is applicable to a broad class of
% distributed physical systems governed by transport phenomena.

% {\color{red}The identified PDE should be interpreted as an effective transport model rather than a unique first-principles description of the FAWG flow. Alternative forcing and boundary-condition formulations may provide other effective representations of the measured input-output behavior. }

\backmatter

\bmhead{Acknowledgements}
This work is supported 
by the National Natural Science Foundation of China (NSFC) 
through grants 
 W2541002, %  NSFC-ICFCRT
and  12572260,  % rotor shape optimization
% and 12172111,  Gao Nan 
% and 12302293   Guy CORNEJO MACEDA
% by the Guangdong Basic and Applied Basic Research Foundation under grant 2022A1515011492,
and by the Shenzhen Science and Technology Innovation Program 
under grants KJZD20230923115210021. % Meituan
We appreciate generous technical and scientific support from the HangHua company (Dalian, China).

The Department of Aerospace Engineering at Universidad Carlos III de Madrid has been accredited as Unit of Excellence ‘María de Maeztu’, a distinction awarded by the Spanish State Research Agency for the period 01/01/2026-31/12/2031.

We sincerely appreciate  advice and support of Profs. Kun Wang, Franz Raps, Yang Jincai, Yang Yang and Chuntai Zheng.

\section*{Declarations}

\subsection*{Funding}
This work is supported 
by the National Natural Science Foundation of China (NSFC) 
through grants 
 W2541002, %  NSFC-ICFCRT
and  12572260,  % rotor shape optimization
% and 12172111,  Gao Nan 
% and 12302293   Guy CORNEJO MACEDA
% by the Guangdong Basic and Applied Basic Research Foundation under grant 2022A1515011492,
and by the Shenzhen Science and Technology Innovation Program 
under grants KJZD20230923115210021. % Meituan
We appreciate generous technical and scientific support from the HangHua company (Dalian, China).

\subsection*{Competing interests}
T. Zhang is the founder and chairman of Ningbo Dongteng Low-Altitude Technology Co., Ltd., and holds equity in the company. The company built the FAWG facility used in this study and provided access to it; no other financial support was received from the company for this work. The other authors have no conflicts to disclose.

\subsection*{Author contributions}
T.~Shaqarin: Conceptualization, Methodology, Software, Formal analysis,
Investigation, Validation, Visualization, Data curation,
 Writing -- original draft,
Writing -- review \& editing.

I.~Al-Manasra: Investigation, Data curation, Software.

T.~Zhang: Resources, Investigation, Writing -- review \& editing.

A.~Ianiro: Formal analysis, Validation,
Writing -- review \& editing.

S.~Discetti: Formal analysis, Validation,
Writing -- review \& editing.

B.~R.~Noack: Resources, Project administration, Conceptualization, Methodology, Supervision,
Writing -- review \& editing.

\subsection*{Data availability}
The data that support the findings of this study are available from the corresponding author upon reasonable request.

\subsection*{Declaration of generative AI in scientific writing}
During the preparation of this work, the authors used large language models in the writing process to improve the manuscript's readability and language. An AI assistant was also used to suggest revisions of the structure and of the analysis, which the authors checked and verified. After using these tools, the authors reviewed and edited the content as needed and assume full responsibility.

\bibliography{shaq2026_AiA}

\end{document}